# RARE KAON DECAYS WITH NA48/2 AND NA62 AT CERN

PATRIZIA CENCI[*]

*INFN Perugia, Via A. Pascoli*
*Perugia, 06123, Italy*
*patrizia.cenci@pg.infn.it*

The Kaon physics program at CERN will be shortly presented, addressing the most recent results from the NA48/2 and NA62 experiments at the CERN SPS and future prospects.

## 1. Introduction

In 2003-04, the NA48/2 experiment has collected, at the CERN SPS, the largest world sample of charged kaon decays, with the main goal of searching for direct CP violation.[1] At an early stage, in 2007-08, the NA62 experiment[2] collected a large minimum bias data sample exploiting the same detector with modified data taking conditions. The main goal was the measurement of the ratio of the rates of the leptonic kaon decays $K^{\pm} \to e\nu_e$ ($K_{e2}$) and $K^{\pm} \to \mu\nu_\mu$ ($K_{\mu 2}$).[3] The large statistics accumulated by both experiments has allowed the studies of a variety of rare kaon decay modes.

## 2. Latest results from the NA48/2 experiment

The beam line of the NA48/2 experiment has been designed to deliver simultaneous, narrow momentum band, $K^+$ and $K^-$ beams produced by 400 GeV/c primary protons extracted from the CERN SPS impinging on a beryllium target. The experimental layout is displayed in Fig. 1 (left). The momentum of the charged particles from $K^{\pm}$ decays was measured by a magnetic spectrometer consisting of four drift chambers (DCH1-DCH4) and a dipole magnet. The spectrometer was located in a tank filled with helium at atmospheric pressure and followed by a scintillator trigger hodoscope. A liquid Krypton calorimeter (LKr) was used to measure the energy of electrons and photons. A hadron calorimeter (HAC) and a muon veto system (MUV), essential to distinguish muons from pions, were located further downstream. The detector description is available in Ref. 4.

---

[*] on behalf of the NA48/2 and NA62 Collaborations at CERN (The NA48/2 Collaboration: Cambridge, CERN, Dubna, Chicago, Edinburgh, Ferrara, Firenze, Mainz, Northwestern University, Perugia, Pisa, Saclay, Siegen, Torino, Wien. The NA62 Collaboration: Birmingham, CERN, Dubna JINR, Fairfax, Ferrara, Firenze, Frascati LNF, Mainz, Merced, Moscow INR, Napoli, Perugia, Pisa, Roma I, Roma Tor Vergata, Saclay IRFU, San Luis Potosi, Stanford, Sofia, Torino, TRIUMF).

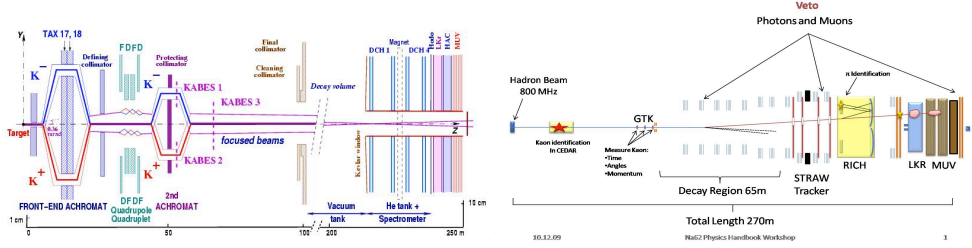

Fig. 1. Schematic drawing of the NA48/2 experimental layout (left) and of the NA62 detector (right).

## 2.1. *The $K^\pm \to \pi^\pm \gamma\gamma$ decays*

Measurements of radiative non-leptonic kaon decays provide crucial tests of the Chiral Perturbation Theory (ChPT), describing weak low energy processes. In ChPT the main contributions to the $K^\pm \to \pi^\pm \gamma\gamma$ decay at the lowest order $O(p^4)$ depend on an unknown parameter ĉ. Higher order corrections including $O(p^6)$ contribution have been found to modify the decay spectrum and rate significantly. A sample of more than 300 $K^\pm \to \pi^\pm \gamma\gamma$ rare decays with a background contamination below 10% has been collected during a 3-day NA48/2 run (2004) and a 3-month NA62 run (2007) at low intensity with minimum bias trigger configuration. The results[5-6] are obtained at increased precision and set stronger constrains on ChPT predictions. The observed decay spectrum agrees with the ChPT description and the preliminary combination of the measured ChPT parameters are $ĉ_4 = 1.56 \pm 0.23$ and $ĉ_6 = 2.00 \pm 0.26$. The preliminary combined branching ratio, calculated in the full kinematic range assuming the $O(p^6)$ description, is $BR_6 = (1.01 \pm 0.06) \times 10^{-6}$. The results agree with the only measurement published so far and represent a significant improvement with respect to the earlier data. The publication of the final combined result is in preparation. The invariant mass distribution of $\pi^\pm\gamma\gamma$ candidates (2007 data), with MC expectation for signal and background, is displayed in Fig. 2 (left).

## 2.2. *Semileptonic kaon decays*

Semileptonic kaon decays offer the most precise determination of the CKM matrix element $|V_{us}|$. The experimental precision is however limited by the knowledge of the form factors (FF) of this decay, since they enter both the phase space integral and the detector acceptances. The NA48/2 experiment presents new measurements of the FF of $K^\pm$ semileptonic decays, based on 4.3 million $K^\pm \to \pi^0 e \nu_e$ ($K^\pm_{e3}$) and 3.5 million $K^\pm \to \pi^0 \mu \nu_e$ ($K^\pm_{\mu3}$) decays collected in 2004, both with negligible background. The hadronic matrix element of the $K^\pm_{\ell3}$ ($\ell = e$ or $\mu$) decays is described by two dimensionless FF which depend on the squared four-momentum transferred to the $\ell$-$\nu$ system. They can be expressed in terms of vector $f_+(t)$ and scalar $f_0(t)$ exchange contributions, parametrized either as a Taylor expansion ("quadratic parametrization"), quantified with $\lambda$ coefficients, or by assuming the dominance of vector (V) or scalar (S) resonance exchange ("pole parametrization"), where pole masses are the only free parameters.

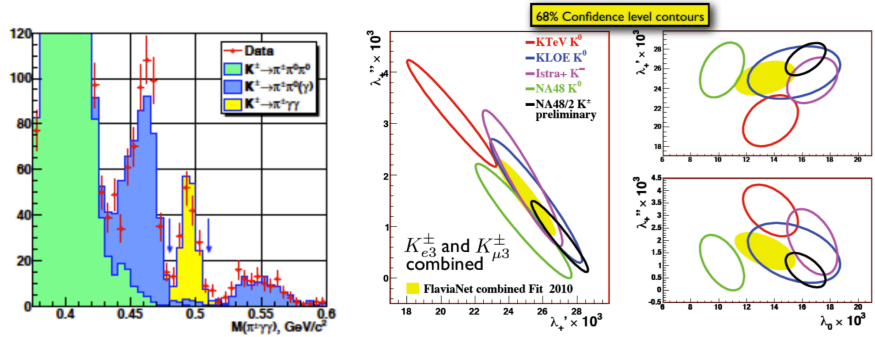

Fig. 2. The $\pi^{\pm}\gamma\gamma$ invariant mass distribution (2007 data) with MC expectation for signal and background (left). Preliminary combined quadratic fit results for $K^{\pm}_{\ell 3}$ decays (right). The ellipses are 68% C.L. contours.

Fig. 2 (right) shows the preliminary comparison of combined quadratic fit results for $K^{\pm}_{\ell 3}$ decays from NA48/2 and other experiments.[7] The ellipses represent 68% Confidence Level (C.L.) contours. The NA48/2 results agree with other experiments and match the precision of the world average on the vector and scalar FF, allowing an improved precision on the $K^{\pm}_{\ell 3}$ form factor contribution to the $|V_{us}|$ uncertainty. The combined $K^{\pm}_{e3}$–$K^{\pm}_{\mu 3}$ results from NA48/2 are the first high precision FF measurement with both $K^+$ and $K^-$ mesons. The comparison of both channels sets tight constraints on lepton flavor violation and other possible new physics.

### 2.3. *The $K^{\pm}\to\pi^+\pi^- e\nu_e$ and the $K^{\pm}\to\pi^0\pi^0 e\nu_e$ decays*

The NA48/2 collaboration has analyzed 1.13 million charged kaon decays $K^{\pm}\to\pi^+\pi^- e\nu_e$ ($K^{+-}_{e4}$) leading to detailed FF studies and an improved determination of the BR at percent level precision.[8] The hadronic FF in the S- and P-wave and their variation with energy are obtained concurrently with the phase difference between the S- and P-wave states of the $\pi^+\pi^-$ system. The latter measurement allows a precise determination of $a_0^0$ and $a_0^2$, the I=0 and I=2 S-wave $\pi^+\pi^-$ scattering lengths. A combination of this result with another NA48/2 measurement, obtained in the study of $K^{\pm}\to\pi^{\pm}\pi^+\pi^-$ decays, brings a further improved determination of $a_0^0$ and the first precise experimental measurement of $a_0^2$. These measurements deliver new inputs to low energy QCD calculations and are crucial tests of ChPT and lattice QCD predictions.

New preliminary results on FF and BR measurements are available for the neutral channel $K^{\pm}\to\pi^0\pi^0 e\nu_e$ ($K^{00}_{e4}$).[9] The NA48/2 experiment collected about 66,000 $K^{00}_{e4}$ decays, increasing the world available statistics by several orders of magnitude. The low background contamination, at 1% level, and the very good $\pi^0$ reconstruction allow the first accurate measurements of FF and of the decay BR. The achieved precision makes possible the observation of small effects such as a deficit of events at low $\pi^0\pi^0$ invariant mass, explained by charge exchange rescattering effects in the $\pi\pi$ system below $2m_{\pi^+}$ threshold. Fig. 3 displays the measurements of both neutral and charged BR based on statistically independent samples collected in 2003-04. The results are: BR($K^{00}_{e4}$) = $(2.585 \pm 0.010_{stat} \pm 0.010_{syst} \pm 0.032_{ext}) \times 10^{-5}$ (preliminary) and BR($K^{+-}_{e4}$) = (4.257 ± $0.004_{stat} \pm 0.016_{syst} \pm 0.031_{ext}) \times 10^{-5}$. The systematic errors include uncertainties on acceptance, resolution, particle identification, beam geometry, trigger efficiencies and radiative corrections. The external error stems from the normalization mode BR uncertainties and is now the dominant error. The hatched band shows the experimental error ($\sigma_{exp}=(\sigma_{stat} \oplus \sigma_{syst})$). The total uncertainty (shaded band) includes the external error.

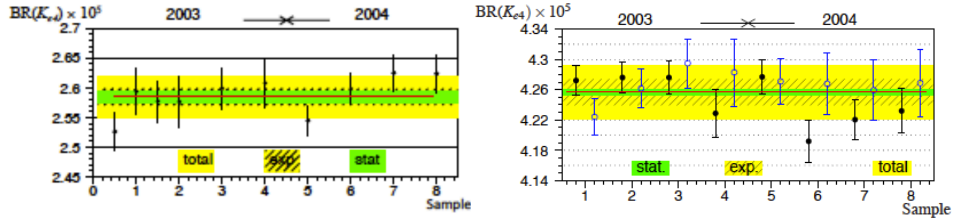

Fig. 3. Measurements of BR($K^{00}_{e4}$) (left) and BR($K^{+-}_{e4}$) (right, both charges) made with different data samples.

## 3. Search for Lepton Flavor Violation and rare decays at the NA62 experiment

The primary goal of NA62[2], the last generation experiment for kaon physics at the CERN SPS, is to measure BR($K^+ \to \pi^+ \nu \bar{\nu}$) with 10% precision. The decay $K^+ \to \pi^+ \nu \bar{\nu}$ is highly suppressed in the Standard Model (SM), while its rate can be predicted with minimal theoretical uncertainty ($BR_{SM} = (0.781 \pm 0.080) \times 10^{-10}$).[10] This BR is a sensitive probe of the flavor sector of the SM. However, its smallness and the challenging experimental signature make it very difficult to measure. The only measurement[11] is compatible with the SM within a large uncertainty. The challenging aspect which drives the design of the NA62 experiment is the suppression of kaon decays with branching fractions of many orders of magnitude higher than the signal This is achieved by collecting about 100 signal events with 10% background in two years of data taking, which requires the observation of at least $10^{13}$ $K^+$ decays in the experiment's fiducial volume. To achieve the needed level of background rejection, NA62 relies on: high-resolution timing, to support a high-rate environment; kinematic rejection, involving cutting on the squared missing mass of the system of incident kaon and observed charged particle under pion assumption; particle identification of kaons, pions, muons, electrons and photons; hermetic vetoing of photons out to large angles from the decay region and of muons within the acceptance; redundancy of information. A suitable experimental apparatus[12] has been conceived to fulfill these purposes. The layout, shown in Fig 1 (right), has the following main features.

- An intense charged hadron beam of secondary particles at 75 GeV/c momentum and 800 MHz average rate, with a 6% $K^+$ component, will be exploited.
- The $K^+$ component in the beam is identified with respect to the other beam particles by an upgraded differential Čerenkov detector (KTAG).
- The coordinates and the momenta of individual beam particles are registered before entering the decay region by 3 silicon pixel tracking detectors (GTK).
- A large-acceptance magnetic spectrometer with low mass tracking detectors in vacuum (STRAW tracker) is required to detect and measure the coordinates and momenta of charged particles originating from the decay region.

- A ring-imaging Čerenkov (RICH) detector, consisting of a long vessel filled with Neon gas, identifies pions with respect to muons.
- A set of photon-veto detectors provides hermetic coverage from zero out to large angles (50 mr) from the decay region. It consists of the existing, high-resolution, electromagnetic calorimeter (LKr), supplemented, at small and forward angles, by intermediate ring (IRC) and small-angle (SAC) Shashlyk calorimeters, and, at large angles, by a series of annular lead-glass photon-veto (LAV) detectors.
- Muon-veto detectors (MUV) are made of a two-part hadron calorimeter followed by iron and a transversally-segmented scintillator hodoscope. This system supplements and provides redundancy with respect to the RICH in muon rejection.
- Scintillator counters (CHANTI) surrounding the last GTK station veto charged particles upstream of the decay region.
- A segmented charged-particle hodoscope (CHOD) made of scintillator counters covers the detector acceptance between the RICH and the LKr calorimeter.
- All the detector components are inter-connected with a high-performance trigger and data-acquisition (TDAQ) integrated system.

Part of the experimental apparatus was commissioned during a technical run in 2012; installation continues and data taking is expected to begin in late 2014.

The features of the NA62 experiment allow a rich physics program, in addition to the main goal. A lepton universality test has been performed using data collected at an early stage of the experiment, in 2007-08. A precise measurement of the helicity-suppressed ratio of the leptonic kaon decay rates, $R_K = \Gamma(K_{e2})/\Gamma(K_{\mu2})$, has been obtained[3]. A world-record sample of about 150,000 reconstructed $K_{e2}$ candidates with 11% background contamination was analyzed. The result $R_K = (2.488 \pm 0.007_{stat} \pm 0.007_{syst}) \times 10^{-5}$ agrees with the SM expectation and is the most precise measurement to date. The record accuracy of 0.4% constrains the parameter space of new physics models with extended Higgs sector, a fourth generation of quarks and leptons or sterile neutrinos.

**References**

1. NA48/2 Collab. (R. Batley *et al*.), Eur. Phys. J **C**52, 875 (2007).
2. NA62 Collab. (G. Anelli *et al*.), CERN-SPSC-2005-013, CERN-SPSC-P-326, (2005).
3. NA62 Collab. (C. Lazzeroni *et al*.), Phys. Lett. **B**719, 326 (2013).
4. NA48/2 Collab. (V. Fanti *et al*.), Nucl. Instrum. Method **A**574, 433 (2007).
5. NA48/2 Collab. (J.R.Batley *et al*.), CERN-PH-EP-2013-197, arXiv:1310.5499 [hep-ex].
6. NA48/2 and NA62 Collab. (E. Goudzovsky *et al*.), *Rare kaon decays*, in Proc. of the European Physical Society Conference on High Energy Physics EPS-HEP2013, PoS(EPS-HEP 2013)426.
7. NA62 Collab. (M. Raggi *et al*.), *High precision measurement of the form factors of the semi leptonic decays at NA48/2,* in Proc. of the 2013 Kaon Physics International Conference, PoS(KAON13)017.
8. NA48/2 Collab. (R. Batley *et al*.), Eur. Phys. J **C**70, 635 (2010), *ibidem* Phys. Lett. **B**715, 105 (2012).
9. NA62 Collab. (B. Bloch-Devaux *et al*.), *Study of the $K^{\pm} \rightarrow \pi^0\pi^0 e \nu_e$ decay with NA48/2 at CERN,* in Proc. of the 2013 Kaon Physics International Conference, PoS(KAON13)028.
10. J. Brod, M. Gorbahn and E. Stamou, *Phys. Rev.* **D**83, 034030 (2011).
11. E949 Collab. (A.V. Artamonov *et al*.), *Phys. Rev.* **D**79, 092004 (2009).
12. NA62 Technical Design, https://cdsweb.cern.ch/record/14049857, NA62-10-07, (2010).